# Electronic and optical properties of novel carbon allotropes

Zhanyu Wang[a], R. J. Zhang[a], Y. X. Zheng[a], L.Y. Chen[a], S.Y. Wang[a,b,c,*], C. Z. Wang[c], K. M. Ho[c], Yuan-Jia Fan[d], Bih-Yaw Jin[d] and Wan-Sheng Su[e,*]

[a]*Shanghai Ultra-Precision Optical Manufacturing Engineering Center and Department of Optical Science and Engineering, Fudan University, Shanghai, 200433, China*

[b]*Key Laboratory for Information Science of Electromagnetic Waves (MoE), Shanghai 200433, China*

[c]*Ames Laboratory, U. S. Department of Energy and Department of Physics and Astronomy, Iowa State University, Ames, Iowa 50011, USA*

[d]*Department of Chemistry, National Taiwan University, Taipei 10617, Taiwan*

[e]*National Center for High-Performance Computing, Hsinchu 30076, Taiwan and Department of Physics, National Chung Hsing University, Taichung 40227, Taiwan*

E-mail: songyouwang@fudan.edu.cn (S. Y. Wang) and wssu@nchc.narl.org.tw (W. S. Su)

## Abstract

The phonon properties, electronic structures and optical properties of novel carbon allotropes, such as monolayer penta-graphene (PG), double-layer PG and T12-carbon, were explored by means of first-principles calculations. Results of phonon calculations demonstrate that these exotic carbon allotropes are dynamically stable. In addition, the bulk T12 phase is an indirect-gap semiconductor having a bandgap of ~4.89 eV. Whereas the bulk material transforms to a 2D phase, the monolayer and double-layer PG become quasi-direct gap semiconductors with smaller band gaps of ~2.64 eV and ~3.27eV, respectively. Furthermore, the partial charge density analysis indicates that the 2D phases retain part of the electronic characteristics of the T12 phase. The linear photon energy-dependent dielectric functions and related optical properties including refractive index, extinction coefficient, absorption spectrum, reflectivity, and energy loss spectrum were also computed and discussed. The structural estimation obtained as well as other findings are in agreement with existing theoretical data. The calculated results are beneficial to the practical applications of these exotic carbon allotropes in optoelectronics and electronics.

## 1. Introduction

Recently, both experimental and theoretical scientists have become enthusiastic about two-dimensional (2D) nanomaterials due to the surge in graphene research[1-7]. That is, graphene opens an avenue to develop the 2D semiconducting materials for multifunctional optoelectronic device applications in the future, and many other 2D materials which exhibit a variety of extraordinary properties are being explored and designed[8-12]. Because of their atomic scale thickness, the existence of quantum confinements, and other unique planar advantages, 2D materials are attractive for use in low-power, smaller, more flexible, and more efficient next-generation nanoelectronic devices, as well as for catalysis, sensing, and energy storage applications[13-18]. The physical and chemical properties of 2D materials are intimately related to their atomic arrangement. However, most of the graphene-like 2D materials share the hexagonal lattice form and in some cases are accompanied by carbon pentagons and heptagons.[1, 19] More new 2D materials comprised of a different topological arrangement of atoms and which are related to novel properties are required to provide the broadest offering available to meet virtually every application.

Following the discovery of graphene, penta-graphene (PG)[1], which is entirely composed of carbon pentagons, was predicted and reported. Such a unique 2D metastable carbon allotrope can be exfoliated from T12-carbon[20], which is a novel 3D carbon allotrope. Thorough and systematic theoretical calculations have been performed to confirm that the monolayer PG (MPG) and T12-carbon (TC) are dynamically and mechanically stable. As depicted in Fig. 1(a), monolayer PG, which consists of a packed layer of 4-coordinated C atoms (C1) sandwiched between two layers of 3-coordinated C atoms (C2), has a thickness of ~1.2 Å. Furthermore, monolayer PG with a large intrinsic quasi-direct bandgap of ~3.48 eV is a desirable candidate for optoelectronics and digital electronics. Equally important, physical properties such as energy gap, bulk modulus, shear modulus, and Vickers hardness of T12-carbon resemble those of diamond phases. Interestingly, the basic characteristics of monolayer PG can be tailored by stacking to form 2D materials, rolling to form 1D nanotubes, and even wrapping to form 0D C20 fullerenes[21].These versatile carbon allotropes are expected to offer opportunities for broad applications in nanoelectronics and optical devices. In order to take full advantage of the properties of these novel materials for eventual technological applications, a better understanding of their electronic structure and optical properties is required.

In this work, we employ first-principles calculations to investigate the structural, electronic, and optical properties of monolayer PG, AB stacked double-layer PG (DPG) and T12-carbon. The paper is organized as follows: in Section 2, the computational techniques adopted in this study are described in detail. Results and discussion of the phonon, structural, electronic, and optical properties are presented in Section 3. Finally, conclusions are given in Section 4.

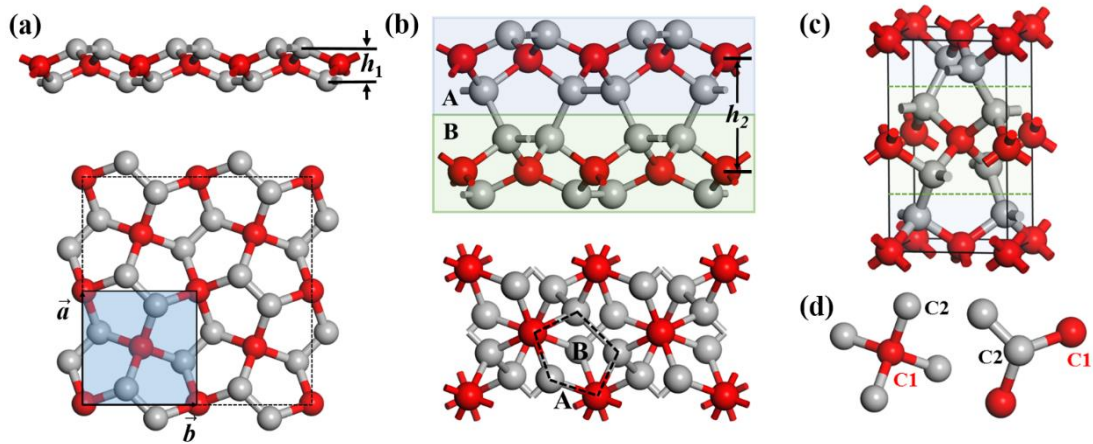

Fig. 1 Side and top views of atomic structure for (a) monolayer PG and (b) AB stacked double-layer PG. Figure 1(a) shows a 2×2 supercell, and the primitive unit cell with lattice vectors is highlighted by a parallelogram. Figure (b) shows that C2 atoms in layer B are located on the hollow sites of layer A. (c) The crystal structure of T12-carbon. (d) Coordination environment of C1 and C2 atoms for monolayer PG.

## 2. Computational details

The static lattice energy and the Hellmann-Feynman force calculations were carried out using the density functional theory (DFT), implemented in the Vienna *ab initio* simulation package (VASP)[22-24]. The projector-augmented wave function (PAW) method[25] with the Perdew-Burke-Ernzerh (PBE)[26] of generalized gradient approximation (GGA) was applied. Such 2D systems are separated by a vacuum distance of ~15 Å in the perpendicular direction. The plane-wave energy cutoff was 500 eV. The Brillouin-zone (BZ) was sampled by 9×9×1 (MPG and DPG) and 9×9×5 (TC) Monkhorst-Pack[27] k meshes for the primitive unit cells. The total energy was calculated with high precision, converged to $10^{-6}$ eV/atom and the lattice constants and the atom coordinates were optimized until the interatomic forces were less than $10^{-2}$ eV/Å.

The phonon calculations were performed using the supercell method through the PHONOPY code[28], and the real-space force constants of supercells were calculated though the density-functional perturbation theory (DFPT) as implemented in VASP. For our phonon calculations, necessary routine numerical checks on the reliability of the supercell size were made. Supercells containing 4×4×1 primitive cells were used for monolayer and double-layer PG, and a 4×4×2 one was used for T12-carbon.

A hybrid HSE06 functional[29, 30] is used for high accuracy electronic structural computations. The complex dielectric function $\varepsilon(\omega)$ is calculated by the so-called independent particle approximation as implemented in VASP.[31] The quasiparticle self-energy corrections, local-field effects, and excitonic contributions are neglected in our calculations.[32] The imaginary part of the dielectric function is obtained from the momentum matrix elements between the occupied and unoccupied wave functions within the selection rules and is given by[33, 34]

$$\varepsilon_2(\omega) = \frac{2e^2\pi}{\Omega\varepsilon_0} \sum_{k,v,c} \left|\left\langle \psi_k^c \left| \hat{u} \times r \right| \psi_k^v \right\rangle\right|^2 \delta\left(E_k^c - E_k^v - E\right), \quad (1)$$

where $\Omega$ is the unit-cell volume, and $\psi_k^c$ and $\psi_k^v$ are the conduction band and valence band wave functions at $k$, respectively. The real part of the dielectric function follows from the Kramer–Kronig relationship[35]. Furthermore, all other optical constants of energy dependence, those of refractive index, extinction coefficient, absorption spectrum, reflectivity, and energy loss spectrum can be derived from $\varepsilon_1(\omega)$ and $\varepsilon_2(\omega)$.

## 3. Results and discussion

### 3.1 Structure and phonon

The optimized crystal structures of k the monolayer PG, AB stacked double-layer PG and T12-carbon are shown in Fig. 1. Monolayer PG belongs to the space group $P\bar{4}2_1m$ with two C1 atoms and four C2 atoms per primitive cell, while the symmetries of AB stacked double-layer PG and T12-carbon are given by space group $Cmme$ and $P4_2/ncm$, respectively. Note that the coordination environment of C2 atoms for double-layer PG and T12-carbon is changed, but for ease of comparison, the same label is still used. The calculated structural parameters for three phases are listed in Table 1. The optimized lattice constants of the monolayer PG and T12-carbon agree quite well with the existing theoretical data[1, 20]. The data show that the lattice constant $a$ of these layered carbon allotropies decreases significantly with an increasing number of atomic layers, while the single layer thickness $h_1$ and layer distance $h_2$ increase considerably. The obvious variations of the structural parameters may be attributed to stacking-induced changes of the interlayer bonding. This investigation demonstrates the evolution of structural parameters in the covalently bonded layered materials in changing from the 3D to the 2D regime.

**Table 1** Calculated lattice constants ($a$ and $c$ in Å), single layer thickness ($h_1$ in Å), layer distance ($h_2$ in Å) and cohesive energy per atom ($E_c/N$ in eV/atom) for monolayer PG, AB stacked double-layer PG and T12-carbon

| Structural | $a$ | $c$ | $h_1$ | $h_2$ | $E_c/N$ |
|---|---|---|---|---|---|
| MPG | 3.644 | - | 1.205 | - | -7.071 |
| DPG | 3.526 | - | 1.493 | 3.037 | -7.354 |
| TC | 3.429 | 6.091 | 1.709 | 3.045 | -7.734 |

Table 1 presents the cohesive energies of the three phases. The cohesive energy of the carbon allotrope is defined as $E_c = E_{total} - N \times E_{C\ atom}$, where $E_{total}$ is the total energy of the allotrope primitive unit cell, $N$ is the number of carbon atoms in the cell and $E_{\text{C atom}}$ is the energy of the isolated carbon atoms. We also calculated the cohesive energies for graphene and diamond for comparison. Considering cohesive energies are found to be -7.973 eV/atom for graphene and -7.841 eV/atom for diamond, the monolayer PG, double-layer PG and T12-carbon are also energetically favorable, though these phases are metastable.

Phonon plays a crucial role in determining the stability of crystal. Figure 2 presents the calculated phonon dispersion curves for monolayer PG, double-layer PG and T12-carbon along the high-symmetry points in the first Brillouin zone (BZ). For monolayer PG, as is a characteristic feature of the phonon dispersion of 2D layered crystals, the out-of-plane acoustical modes (ZA) display parabolic dispersion because the transverse forces decay exponentially, in contrast to the in-plane longitudinal acoustic (LA) and transverse acoustic branches (TA) showing linear dispersions near the Γ point. However, this feature for double-layer PG is not as obvious as that for monolayer PG. From the partial phonon density of states (PDOS) in Fig. 2, the highest-frequency optical phonons at ~ 48 THz dominated by the vibrations of $sp^2$-hybridized C2 atoms in monolayer and AB stacked double-layer PG disappear in fully $sp^3$-hybridized T12-carbon. This indicates that the strong interlayer interaction in T12-carbon suppresses C2 atom vibrations. Most importantly, phonon calculations indicate that these phases are dynamically stable, because no imaginary frequencies in the Brillouin zone are found for any of them.

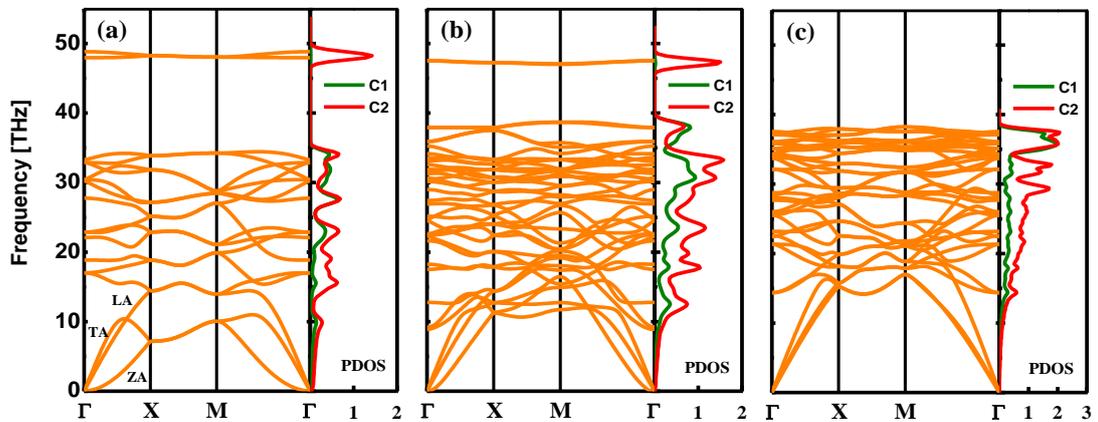

**Fig. 2** *Ab initio* calculation of phonon dispersion curves along the Γ-X-M-Γ directions

of the BZ and vibrational partial density of states (PDOS) for monolayer PG, AB stacked double-layer PG and T12-carbon at the equilibrium volume.

**3.2 Electronic properties**

Changes in interlayer interaction and degree of quantum confinement not only result in significant variations of the structural parameters and vibrational properties, but also lead to dramatic differences in the electronic structure of monolayer and double-layer PG compared with their bulk counterpart. This effect has been particularly demonstrated by semiconducting $MoS_2$.[36] However, compared to the strong covalent bonding between adjacent layers of double-layer PG and T12-carbon, the bulk $MoS_2$ differs in that the interlayer interactions along the crystallographic $c$ axis consist mainly of weak van der Waals forces, stemming from the chemically saturated chalcogen atoms. In order to analyze the electronic properties of 2D monolayer and double-layer PG as well as bulk T12-carbon, their energy band structures are illustrated in Fig. 3(a)-(d). The Fermi level is shifted to 0 eV. The bulk T12 phase is an indirect-gap semiconductor having a bandgap of ~4.89 eV with a valence band maximum (VBM) at the Γ point and a conduction band minimum (CBM) at the point along the A-Z symmetry line. As the bulk material transforms to 2D phase, the band gap decreases to ~2.64 eV for double-layer PG and ~3.27 eV for monolayer PG. The double-layer and monolayer PG are also indirect-gap semiconductors whereas the top branches of the valence bands for these 2D phases are very flat. For instance, the value of the valence band at the VB1 point for monolayer PG is very close to that of VBM. Hence, these 2D phases, especially for the monolayer PG, can be viewed as quasi-direct gap semiconductors.

Intriguingly, the overall features of band structures for the three phases appear quite similar, as can be seen in Fig. 3(a)-(c), except that there are four new bands for the 2D phases located in the original bandgap of T12-carbon. The indirect-to-quasi-direct bandgap transition from bulk to 2D material arises from the four bands in the original bandgap. Consequently, the bandgap becomes much smaller than the bulk material. The transition is manifested in partial charge densities of some selected points near the Fermi level of the three phrases shown in Figs. 3(e)-(g). For T12-carbon, VBM's electronic densities are seen to be primarily localized along the bonds between the C1 and C2 atoms in a single layer, while CBM's electronic densities are mainly localized between two covalently bonded layers. Additionally, the partial charge densities of CB1 appear very similar to that of CBM and are not shown in the figure. Meanwhile, it is worth noting that the partial charge densities of VB1 (CB1) for double-layer PG are also very similar to that of VBM (CBM) in the T12 phase. That is, double-layer PG inherits a portion of the electronic properties from the T12 phase. For monolayer PG, the distribution of partial charge densities for VBM and CBM is dramatically affected by the quantum confinement effect, and is visibly different from that of double-layer PG. However, there are only minor differences between the band structures of the monolayer and double-layer PG.

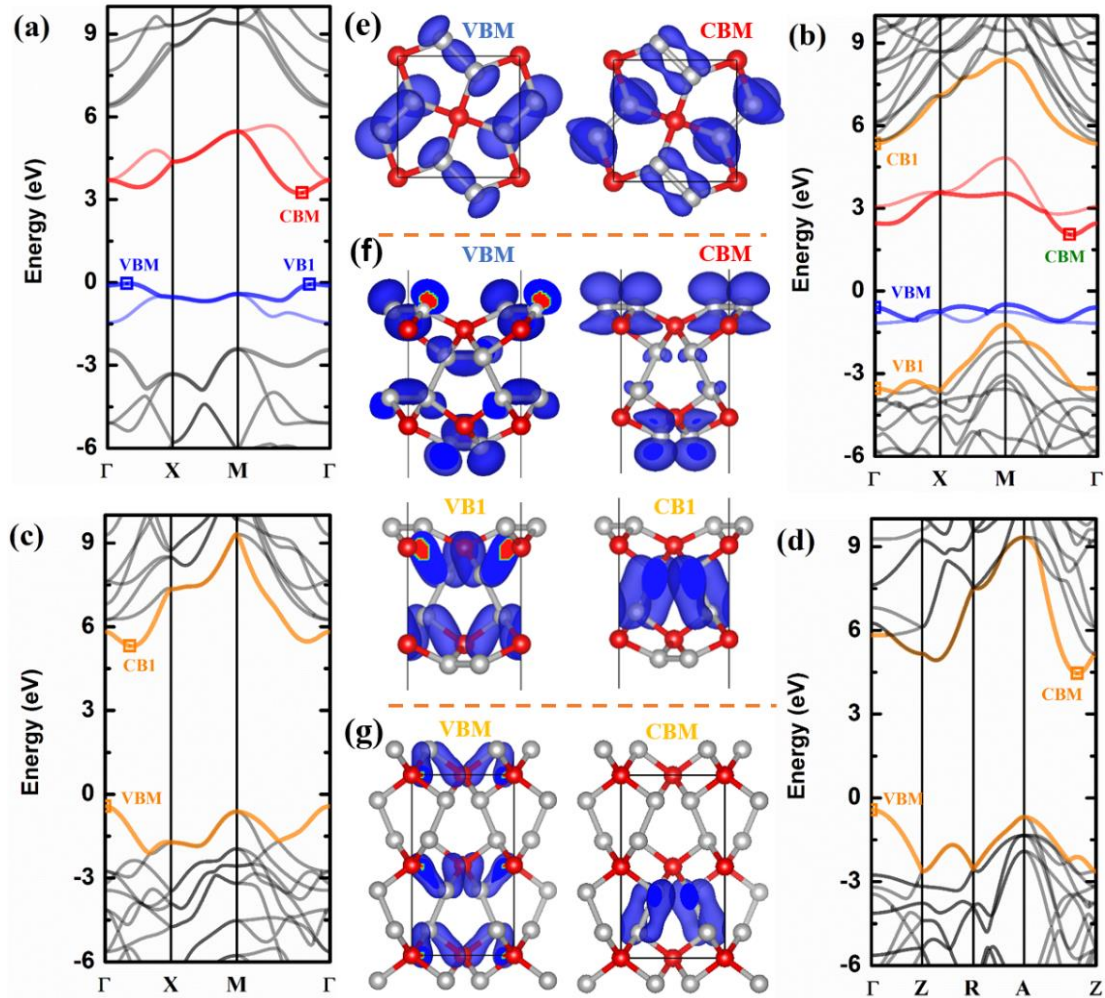

**Fig. 3** Calculated band structures of (a) monolayer PG, (b) double layer PG and (c) T12-carbon along the Γ-X-M-Γ directions of the BZ. (d) Band structure of T12-carbon along the Γ-Z-R-A-Z. Partial charge densities of specific states for (e) monolayer PG, (f) double layer PG and (g) T12-carbon are also shown.

To gain further insight into corresponding electronic structures, the total and partial density of states of monolayer PG, double-layer PG and T12-carbon are presented in Fig. 4. The conduction and valence bands of T12-carbon are composed mostly of C1 2$p$ and C2 2$p$ states. For monolayer and double-layer PG, the lowest conduction bands are mainly attributed to C2 2$p$ states, and the top of the valence bands are dominated by hybridizing C2 2$p$ and C1 2$p$ states.

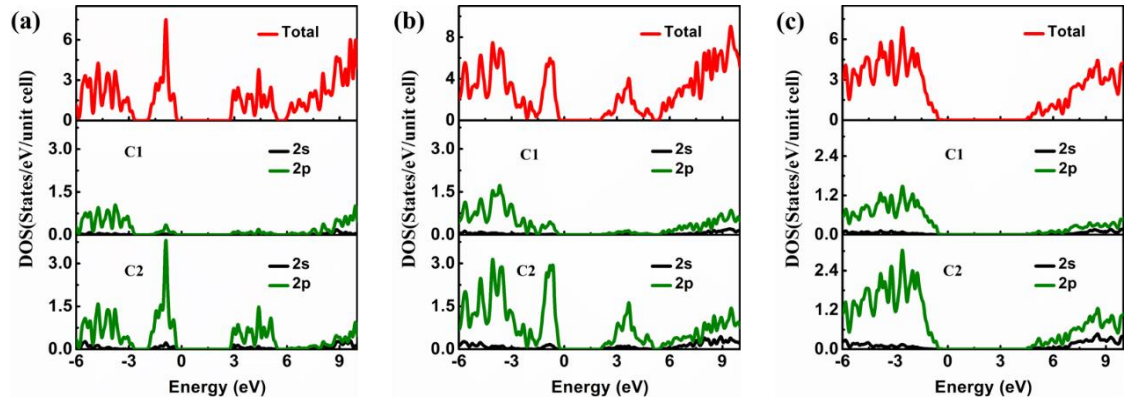

**Fig. 4** Calculated partial and total densities of states diagrams for (a) monolayer PG, (b) double layer PG, and (c) T12-carbon.

It is well known that in graphene, the $s$, $p_x$, and $p_y$ atomic orbitals on each carbon hybridize to form strong covalent $sp^2$ bonds, and the remaining $p_z$ orbital overlaps with three neighboring carbons to form long-range π-conjugation states, leading to the gapless nature of graphene. Originating from the full $sp^3$ bonding feature in the T12-carbon, all the $s$, $p_x$, $p_y$ and $p_z$ atomic orbitals are localized on the $sp^3$-hybridized C atoms. Thus, the localized states will open a wide bandgap in the T12 phase. Interestingly, monolayer and double-layer PG show a mixed character of graphene and T12-carbon; that is, they exhibit $sp^2$-$sp^3$ hybrid carbon structures. Combining the partial charge densities and density of states, the VBM and CBM of monolayer and double-layer PG are dominated by the $p_z$ states of $sp^2$-hybridized C2 atoms. However, these delocalized states in both 2D structures are spatially separated by the $sp^3$-hybridized C1 atoms and full electron delocalization is hindered. Accordingly, monolayer and double-layer PG have bandgaps somewhere between graphene and T12-carbon.

### 3.3 Optical properties

Here, optical properties of monolayer PG, double-layer PG and T12-carbon are calculated and discussed. First, for the three phases, the imaginary $\varepsilon_2(\omega)$ and real $\varepsilon_1(\omega)$ parts of the dielectric functions as a function of the photon energy for both $E$ vectors perpendicular and parallel to the $c$ axis are illustrated in Fig. 5. For double-layer PG, the dielectric functions are found to be highly anisotropic in the low energy range (<12 eV) and to become isotropic in the high energy range. However, for T12-carbon, it is clear that the behavior of $\varepsilon_2(\omega)$ and $\varepsilon_1(\omega)$ for $E\perp c$ and $E\perp c$ are very similar; that is, its optical properties show no sizable anisotropy. Furthermore, it is worth mentioning that the effect of quantum confinement introduces four new states and shortens the band gap in the 2D phases, so for monolayer and double-layer PG the line shapes of $\varepsilon(\omega)$ below ∼6 eV, which is close to the band gap of T12-carbon, are significantly altered. For higher energy ranges, however, the structure of the dielectric functions of monolayer and double-layer PG are found to be similar to that of bulk T12-carbon. This is because 2D PG inherits the main band structure features from the T12 phase.

The origins of the different peaks in the calculated imaginary part of the dielectric function $\varepsilon_2(\omega)$ can be attributed to interband transitions. It is worth mentioning that a particular structural peak in $\varepsilon_2(\omega)$ may correspond to multiple interband transitions since many quasi-direct or indirect transitions may be found in the band structure with an energy corresponding to the same peak. For example, the sharp peaks below ∼ 6 eV for the monolayer and double-layer PG can be mainly assigned to transitions from two valence bands of the four new bands to two conduction bands. According to the analysis of density of states, all the interband transitions for the three phases are mainly due to transitions from $2p$ of C2 to the $2p$ of C2. The contributions from transitions from $2p$ of C1 to the $2p$ of C2 as well as $2p$ of C2 to the $2p$ of C1 are also very large for the peak above ~ 6 eV. Considering the selection rules, only transitions with the difference $\Delta l = \pm 1$ between the angular momentum quantum numbers $l$ are allowed, i.e., the atomic $p$-$p$ transition is forbidden. However, in the three phrases, due to the orbital hybridization, the valence and conduction bands still have $2s$ orbital contributions; thus the transition dominated by the $2p$-$2p$ transition is still allowed.

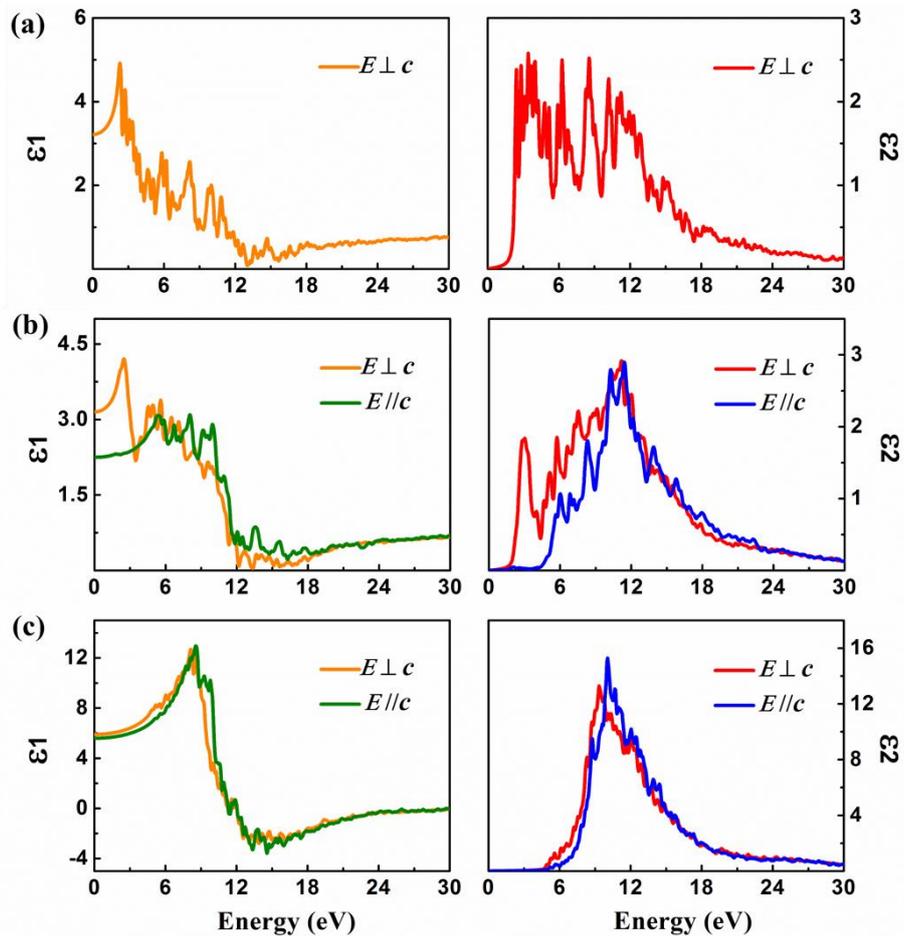

**Fig. 5** Calculated real $\varepsilon_1(\omega)$ and imaginary $\varepsilon_2(\omega)$ parts of the dielectric function versus photon energy of (a) the monolayer PG, (b) double layer PG and (c) T12.

Figure 5(a)–(c) depicts the calculated results for refractive index, extinction

coefficient, absorption spectrum, reflectivity, and energy loss spectrum for monolayer PG, double-layer PG and T12-carbon, respectively. For the $E\perp c$ of monolayer PG, double-layer PG and T12-carbon, the static refractive index is found to have a value of 1.79, 1.78 and 2.43, respectively. The refractive index n exhibits oscillatory variation below ~18 eV for monolayer and double-layer PG. For T12-carbon, the refractive index reaches a maximum value of 3.71 at 8.64 eV. Strong absorption is found in the 7–40 eV range with a very broad peak for T12-carbon. For the 2D phases, strong absorption is also found in the 6–24 eV range, and there are also additional peaks below 6 eV. For each phase, there is no absorption and only a small reflectivity in the photon-energy range up to an energy value close to the corresponding bandgap, meaning that the material becomes transparent in this region. The energy-loss spectrum describes the energy loss of a fast electron traversing a material. The main peak is located at ~ 15.8 eV, 18.1 eV and 32.7 eV for the $E\perp c$ of monolayer PG, double-layer PG and T12-carbon, respectively. Each of these values is generally defined as the plasma frequency, and corresponds to the trailing edges in the reflection spectra.

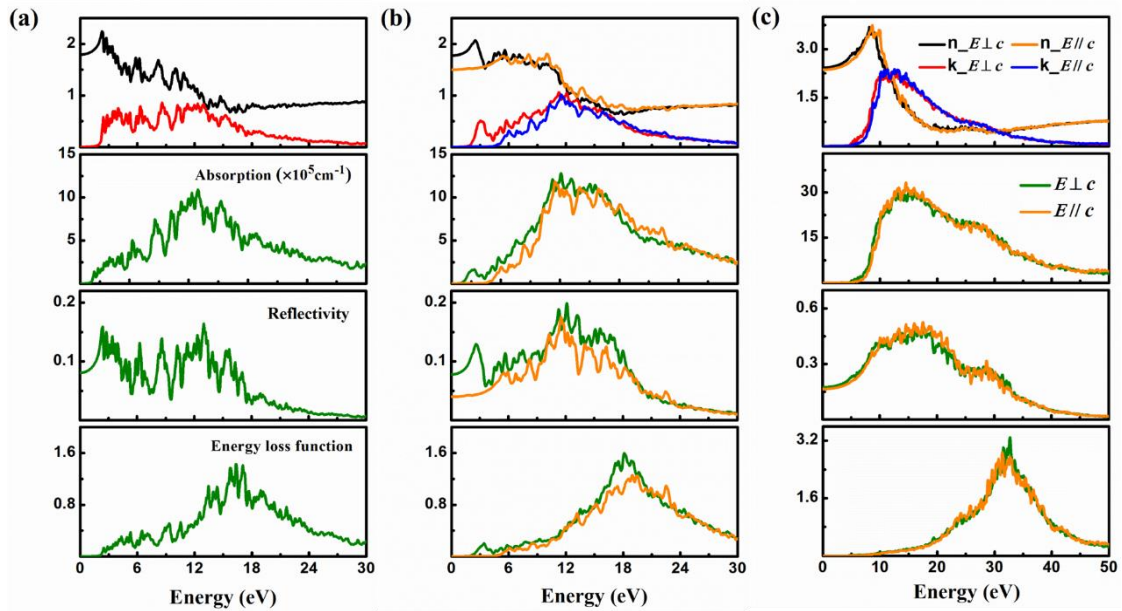

**Fig. 6** Calculated refractive index n, extinction coefficient k, absorption spectrum, reflectivity, and energy-loss spectrum for (a) monolayer PG, (b) double layer PG and (c) T12.

## 4. Conclusions

In summary, first principles calculations have been performed to obtain the structure as well as the phonon spectra, electronic and optical properties of monolayer PG, AB stacked double-layer PG and T12-carbon. The optimized lattice constants of the monolayer PG and T12-carbon agree quite well with the available theoretical data. These carbon allotropes have slightly higher cohesive energies than graphene and diamond, and are also energetically favorable. Phonon calculations indicate that the three phases are dynamically stable. The bulk T12 phase is an indirect-gap semiconductor having a bandgap of ~4.89 eV, whereas the double-layer and monolayer PG become quasi-direct gap semiconductors with the band gap decreasing to ~2.64 eV for double-layer PG and ~3.27 eV for monolayer PG. The indirect-to-quasi-direct bandgap transition arises from the four new bands in the original bandgap. Meanwhile, monolayer and double-layer PG inherit the main band structure features from the T12 phase. Finally, the linear photon energy dependent complex dielectric functions and related optical properties including refractive index, extinction coefficient, absorption spectrum, reflectivity, and energy loss spectrum were computed and discussed. Our investigations are beneficial to the practical applications of these exotic carbon allotropes in optoelectronics and electronics.

## Acknowledgments

W. S. Su would like to thank the Ministry of Science and Technology for financially supporting this research under Contract No. MOST-104-2112-M-492-001. Support from the National Centers for Theoretical Sciences and High-performance Computing of Taiwan in providing significant computing resources to facilitate this research are also gratefully acknowledged. Work at Fudan University was supported by the NSF of China (Grant No. 11374055 and 61427815), and National Basic Research Program of China (No. 2012CB934303 and 2010CB933703). Work at Ames Laboratory was supported by the US Department of Energy, Basic Energy Sciences, and Division of Materials Science and Engineering, including a grant of computer time at the National